\documentclass[12pt]{spieman}  
\usepackage{amsmath,amsfonts,amssymb}
\usepackage{graphicx}
\usepackage{setspace}
\usepackage{tocloft}

\newcommand{\rev}[1]{#1} 

\title{Design, pointing control, and on-sky performance of the mid-infrared vortex coronagraph for the VLT/NEAR experiment}

\author[a,*]{Anne-Lise Maire}
\author[b]{Elsa Huby}
\author[a]{Olivier Absil}
\author[c]{G\'erard Zins}
\author[d]{Markus Kasper}
\author[a]{Christian Delacroix}
\author[d]{Serban Leveratto}
\author[e]{Mikael Karlsson}
\author[f,g]{Garreth Ruane}
\author[d]{Hans-Ulrich K\"aufl}
\author[a]{Gilles Orban de Xivry}
\author[d]{Prashant Pathak}
\author[d]{Lorenzo Pettazzi}
\author[d]{Philippe Duhoux}
\author[c]{Johan Kolb}
\author[h]{\'Eric Pantin}
\author[g]{A.~J.~Eldorado Riggs}
\author[d]{Ralf Siebenmorgen}
\author[f,g]{Dimitri Mawet}
\affil[a]{STAR Institute, Universit\'e de Li\`ege, All\'ee du Six Ao\^ut 19c, B-4000 Li\`ege, Belgium}
\affil[b]{LESIA, Observatoire de Paris, Universit\'e PSL, CNRS, Sorbonne Universit\'e, Universit\'e de Paris, 5 place Jules Janssen, 92195 Meudon, France}
\affil[c]{European Southern Observatory, Alonso de Cordova 3107, Casilla 19001 Vitacura, Santiago 19, Chile}
\affil[d]{European Southern Observatory, Karl-Schwarzschild-Str. 2, 85748 Garching bei M\"unchen, Germany}
\affil[e]{Department of Materials Science and Engineering, Angstr\"om Laboratory, Uppsala University, PO Box 534, 751\,21 Uppsala, Sweden}
\affil[f]{Department of Astronomy, California Institute of Technology, Pasadena, CA 91125, USA}
\affil[g]{Jet Propulsion Laboratory, California Institute of Technology, Pasadena, CA 91109, USA}
\affil[h]{Laboratoire CEA, IRFU/DAp, AIM, Universit\'e Paris-Saclay, Universit\'e Paris Diderot, Sorbonne Paris Cit\'e, CNRS, 91191, Gif-sur-Yvette, France}

\cftpagenumbersoff{figure}
\cftpagenumbersoff{table} 
\begin{document} 
\maketitle

\begin{abstract}
Vortex coronagraphs have been shown to be a promising avenue for high-contrast imaging in the close-in environment of stars at thermal infrared (IR) wavelengths{. They} are included in the baseline design of the \rev{mid-infrared extremely large telescope imager and spectrograph}. To ensure good performance of these coronagraphs, a precise control of the centering of the star image in real time is needed. We previously developed and validated the \rev{quadrant analysis of coronagraphic images for tip-tilt sensing estimator (QACITS)} pointing {estimator} to address this issue. {While this approach is not wavelength-dependent in theory, it} was never {implemented for mid-IR observations, which leads to specific challenges and limitations. Here,} we present the design of the mid-IR vortex coronagraph for the new Earths in the $\alpha$ Cen Region (NEAR) experiment with the Very Large Telescope (VLT)/Very Large Telescope imager and spectrometer for the mid-infrared (VISIR) instrument and assess the performance of the QACITS estimator for the centering control of the star image onto the vortex coronagraph. We use simulated data and on-sky data obtained with VLT/VISIR, which was recently upgraded for observations assisted by adaptive optics in the context of the NEAR experiment. We demonstrate that {the QACITS-based correction loop} is able to control the centering of the star image onto the NEAR vortex coronagraph with a stability down to $0.015 \lambda/D$ rms over 4~h in good conditions. These results show that QACITS is a robust approach for precisely controlling in real time the centering of vortex coronagraphs {for mid-IR observations}.
\end{abstract}

\keywords{adaptive optics; infrared imaging; data processing; simulations}

{\noindent \footnotesize\textbf{*}Anne-Lise Maire,  \linkable{almaire@uliege.be} }

\begin{spacing}{2}   

\section{Introduction}

Imaging rocky planets in the habitable zone of nearby stars is one of the major goals of the extremely large telescope (ELTs) for the next decade. Before high-contrast imaging instruments come online on the ELTs, the 100-h observing campaign referred to as \rev{``new Earths in the $\alpha$ Cen region}'' (NEAR)\cite{Kasper2017, Kasper2019} searched for massive rocky exoplanets in the habitable zone around $\alpha$ Centauri A and B with the Very Large Telescope imager and spectrometer for the mid-infrared (VISIR)\cite{Lagage2004} in May to June 2019. The project was funded by the Breakthrough Initiatives and \rev{the European Southern Observatory (ESO)}. The motivation of the project was twofold. The first objective was to image rocky exoplanets in the habitable zone of the nearest stars to the Sun. This science goal was partly motivated by the discovery of a rocky exoplanet in the habitable zone of Proxima Centauri ($\sim$0.05~au)\cite{AngladaEscude2016}, which was followed more recently by the discovery of a potential second planet on a wider orbit ($\sim$1.5~au)\cite{Damasso2020}. To achieve the challenging science goal of the NEAR project, VISIR was upgraded for high-contrast imaging and mounted on the VLT Unit Telescope 4 to use it with the adaptive optics facility (AOF)\cite{Arsenault2017}. In addition, new vortex coronagraphs were installed in VISIR. {As demonstrated during the commissioning and scientific operations of the vortex coronagraph on Keck/NIRC2\cite{Huby2017}, as well as at the \rev{Large Binocular Telescope} and VLT\cite{Absil2016}, a crucial aspect to the performance, efficiency, and data quality of this type of coronagraphic observations is the automatic centering and pointing control of the star onto the vortex phase mask. Based on our previous experience, we, therefore, decided to put a significant effort into this aspect to maximize the scientific return of NEAR.}

A secondary objective of {the NEAR experiment} was to serve as a pathfinder experiment for the \rev{mid-infrared ELT imager and spectrograph (METIS)} \cite{Brandl2014}. Exoplanet imaging is one of the main science drivers for METIS, which includes dedicated high-contrast imaging modes. They rely on the use of adaptive optics (AO), advanced coronagraphy including vortex coronagraphy, and pupil tracking to exploit angular differential imaging (ADI)\cite{Marois2006a}. These techniques were poorly tested on sky at mid-infrared (mid-IR) wavelengths before NEAR, {and never combined with AO in this wavelength range, which provides a yet unexplored level of wavefront correction (Strehl ratio $>$95\%)}. Another important part of the observing strategy at mid-IR wavelengths is the calibration of the high and fluctuating thermal background, associated with sky and instrumental emissivity. The standard strategy based on chopping and nodding {is anticipated to be complex to implement for the ELT and would probably suffer from low efficiency}. Thus, it was interesting to investigate with NEAR the performance of a strategy based on chopping only. The NEAR commissioning\cite{Kasper2019} demonstrated sensitivities after image chopping in the background-limited regime three times superior with respect to the sensitivities achieved with VISIR without AO. {Using the NEAR data, it was shown that the scaling of the achieved signal-to-noise ratio (S/N) follows the square root of the observing time\cite{Kasper2019}.}

In this paper, we validate and analyze the on-sky performance of the new vortex coronagraph used for the NEAR campaign, and its dedicated pointing control system. First, we briefly present the design and the laboratory tests of the NEAR vortex coronagraphs (Sec.~\ref{sec:agpm}). Then, we describe the quadrant analysis of coronagraphic images for tip-tilt sensing (QACITS) approach used for the precise pointing control of the star image on the vortex coronagraph in real time and the expected performance from simulations (Sec.~\ref{sec:qacitsnear}). We present in Sec.~\ref{sec:onskyresults} the results of the on-sky tests to validate the pointing control, and the performance of the coronagraph in terms of starlight rejection. {To our knowledge, this is the first analysis reporting the on-sky performance of a vortex coronagraph for mid-IR observations.}

\section{The VLT/NEAR vortex coronagraph}
\label{sec:agpm}

    \subsection{Mid-infrared annular groove phase masks}
    \label{sec:agpm_masks}

The \rev{annular groove phase mask} (AGPM)\cite{Mawet2005} is an implementation of a vortex phase mask, where the phase ramp is produced by the form birefringence of a circularly symmetric subwavelength grating (i.e., a grating with a period smaller than the operating wavelength). Over the last decade, several AGPMs have been manufactured on synthetic diamond substrates for applications in the thermal IR, from 2 to 13~$\mu$m\cite{Absil2016}. Among these, a small series of N-band AGPMs were specifically developed in 2012 for VISIR\cite{Delacroix2012}. Due to the lack of testing facility at that time, the performance of these AGPMs could not be tested before installation. We, therefore, selected the component that had the best theoretical performance based on our scanning electron microscopy assessment of the grating parameters, leading to the installation of a phase mask referred to as AGPM-N4 in VISIR, while another one (AGPM-N3) was preserved in our lab for potential future testing.

Due to the uncertainties in the actual performances of AGPM-N3 and N4, to their slightly inappropriate optimization bandwidth (11--13.2~$\mu$m) for the NEAR filter (10--12.5~$\mu$m), and to the need for additional AGPMs for the Breakthrough Initiatives, \rev{three new AGPMs were designed and etched. The rejection performances of these AGPMs and of the first-generation AGPM-N3 mask were measured on a thermal IR bench at ESO Garching\cite{Kaeufl2018}. The measured rejection rate of the new AGPMs ($\lesssim 200$ at 10.2~$\mu$m) was found to be inferior to the measured performance of the AGPM-N3 mask ($\sim 400$ at 10.2~$\mu$m).} Based on these performance tests, the best AGPM mask among the new devices (referred to as AGPM-BT3) and the AGPM-N3 mask were sent to Paranal for integration in VISIR alongside the AGPM-N4 mask, whose intrinsic rejection rate was still mostly unknown. Based on tests with the VISIR internal source, its rejection rate was estimated to be $>$100.

    \subsection{Apodized Lyot stops}
    \label{sec:apodlyotstops}

The first-generation AGPMs installed in VISIR in 2012 were combined with conventional Lyot stops, glued on filters. A conventional Lyot stop is typically similar in shape to the telescope pupil, but with a smaller outer diameter and oversized obscurations. These Lyot stops were not specifically optimized for throughput nor rejection, in part because the rejection was expected to be limited by the poor wavefront quality.

\begin{figure}[t]
\centering
\includegraphics[width=.55\textwidth, trim = 0mm 0mm 0mm 0mm, clip]{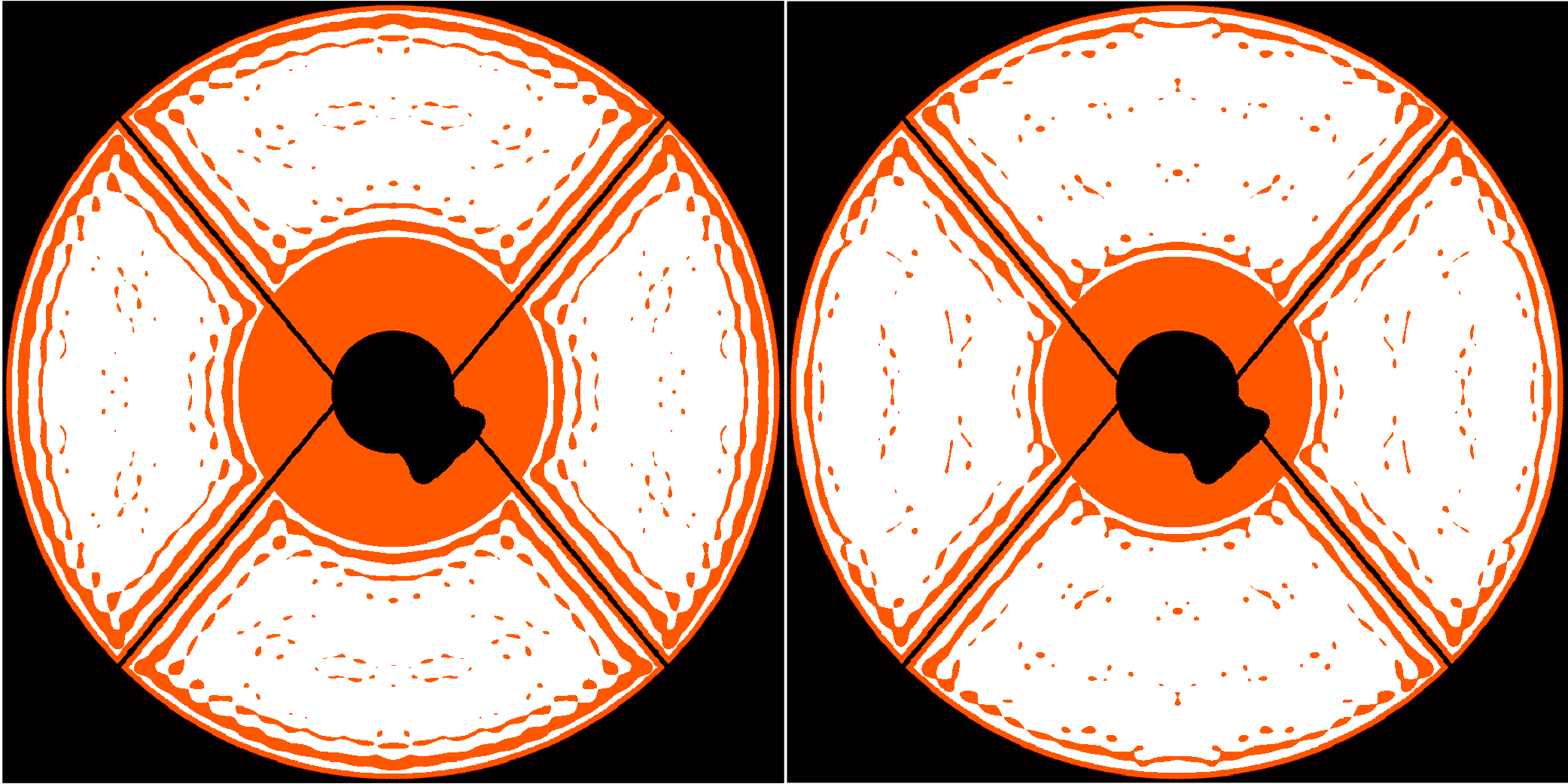}
\caption{Apodized Lyot stops for the {most} aggressive configuration (\textit{left}) and the {least} aggressive configuration (\textit{right}). The transmitting areas of the Lyot stop {are shown in white and the nontransmitting areas of the Lyot stop in orange}. The central obscuration consists of a round shape due to the M2 mirror and a bump to the lower right due to the M3 mirror {(black)}. \rev{The throughputs within a circular aperture of 1~$\lambda/D$ centered on the core of the PSF amount to 37\% of the core throughput of the VLT pupil for the most aggressive configuration and 50\% for the least aggressive configuration.} The NEAR campaign used the Lyot stop on the right.}
\label{fig:apodstops}
\end{figure}

New Lyot stops were specifically optimized for the NEAR campaign (see Fig.~\ref{fig:apodstops}). These Lyot stops were designed to maximize the S/N for planets within 2$''$ of the on-axis star. While the AGPM diffracts light from the on-axis star outside of the Lyot stop, a custom binary apodization pattern is used in the Lyot stop to reduce the diffraction pattern for angular separations from 3$''$ to 8$''$, in order to reduce the diffracted light from the bright off-axis stellar companion within the habitable zone around the on-axis star. The apodization pattern was optimized following Refs.~\citenum{Carlotti2011} and \citenum{Vanderbei2012}. Specifically, our Lyot stop optimization code uses the Gurobi solver\cite{gurobi} to maximize the total energy throughput of the Lyot stop under constraints on the contrast in the predefined region of the focal plane. 

\begin{figure*}[t]
\centering
\includegraphics[width=.8\textwidth, trim = 0mm 0mm 0mm 0mm, clip]{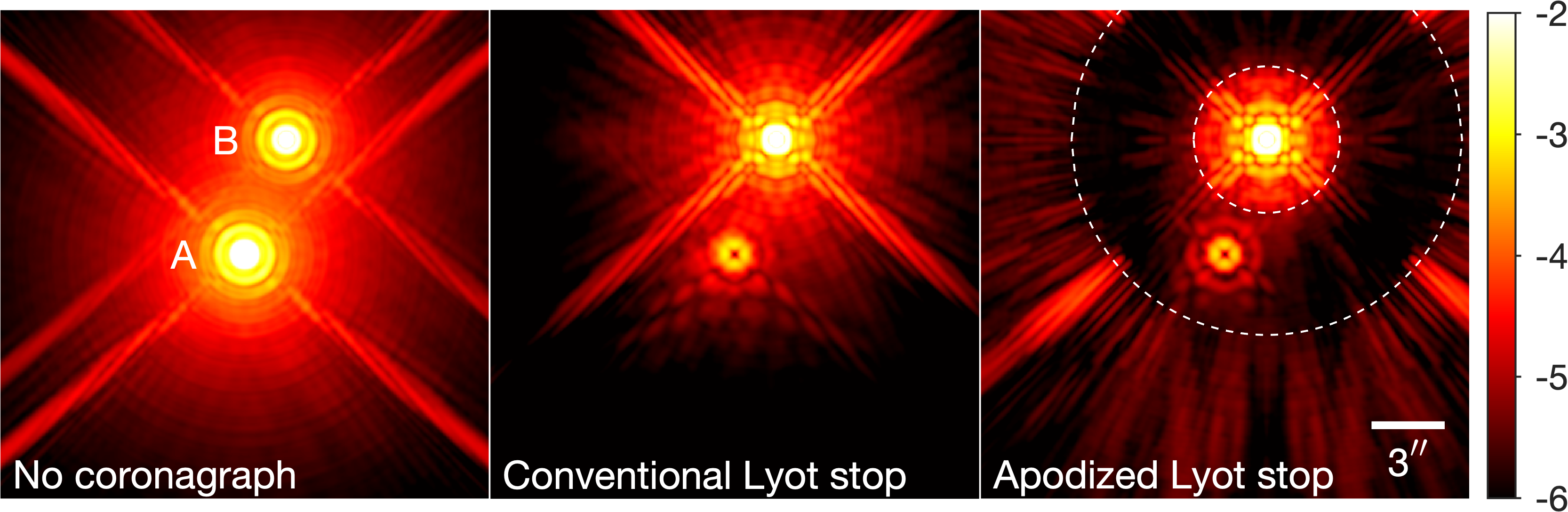}
\caption{Simulated VISIR images of $\alpha$ Cen without a coronagraph (\textit{left}), with an AGPM and a conventional Lyot stop that masks only the M2 and M3 mirrors as well as the telescope spiders (\textit{middle}), and with an AGPM and the apodized Lyot stop used for the NEAR campaign (\textit{right}, see Fig.~\ref{fig:apodstops}). In the \textit{middle} and \textit{right} panels, $\alpha$~Cen~A is suppressed by the coronagraph. In the \textit{right} panel, the white dashed circles indicate the apodized region around $\alpha$~Cen~B. The images were simulated using 29 discrete spectral channels covering the NEAR bandwidth (10--12.5~$\mu$m) assuming a flat source spectrum for both stars. {Even though the choice of a flat spectrum for the stars may not be a good approximation of the Rayleigh-Jeans regime for a blackbody, it is not expected to significantly affect the optimization.} The separation and the flux ratio between the two stars were assumed to be 5$''$ and 0.5, respectively. The color scale on the right shows the intensity in logarithmic scale.}
\label{fig:psfbin}
\end{figure*}

Two designs were produced and manufactured for the NEAR campaign: (i)~an aggressive design and (ii)~a relaxed design. The aggressive design apodizes the off-axis star to give a theoretical raw contrast of $1\times10^{-6}$ between 10 and 29 $\lambda/D$\rev{, where $\lambda$ is the observing wavelength and $D$ the telescope diameter,} assuming there is no wavefront error. The {throughput within a circular aperture of 1~$\lambda/D$ centered on the core of the point-spread function (PSF) amounts to} 37\% of the core throughput of the VLT pupil. The diameter of the Lyot stop's central obscuration is 40\% of the telescope diameter. The relaxed design provides a theoretical raw contrast below $5\times10^{-6}$ between 11 and 29~$\lambda/D$ with a relative core throughput of 50\%, and the central obscuration diameter is 35\% of the telescope diameter. In each case, the outer diameter of the apodized Lyot stops is slightly undersized with respect to the telescope pupil (by a few \%) in order to mitigate the effect of pupil drifts while limiting the throughput loss. The effect of the apodization on the coronagraphic image is shown in Fig.~\ref{fig:psfbin}. The manufactured devices were directly glued onto the NEAR filters and were installed in VISIR during the NEAR upgrade. The relaxed, higher-throughput design was used for the NEAR campaign because, after an initial analysis of the expected wavefront error and the noise in the system, our team concluded that it was likely to provide the best S/N on the $\alpha$~Cen system.

\begin{figure}[t]
\centering
\includegraphics[width=.6\textwidth, trim = 0mm 0mm 0mm 0mm, clip]{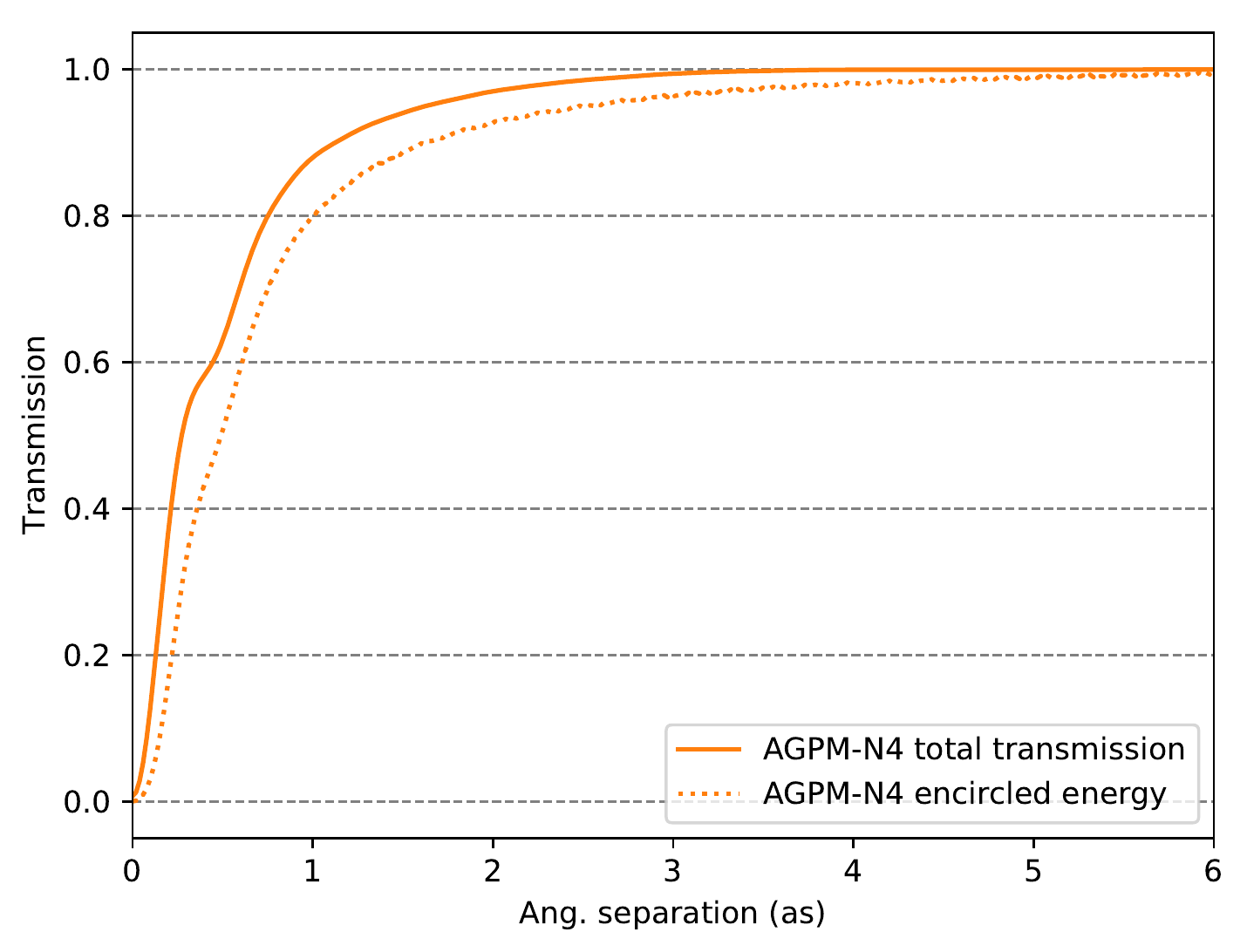}
\caption{\rev{Transmission of the AGPM-N4 mask as a function of the angular separation computed with respect to the case without the AGPM-N4 mask but with the Lyot stop for both the total energy and the energy encircled within 1~$\lambda/D$.}}
\label{fig:agpm_trans}
\end{figure}

\rev{We show in Fig.~\ref{fig:agpm_trans} the radial profile of the transmission of the AGPM-N4 mask computed with respect to the case without the AGPM-N4 mask but with the Lyot stop for both the total energy and the energy encircled within 1~$\lambda/D$.}

\section{Vortex centering with QACITS for NEAR}
\label{sec:qacitsnear}

\subsection{QACITS model adapted for NEAR}

The performance of vortex coronagraphs in terms of starlight rejection is {very} sensitive to a suboptimal centering of the star image behind the mask, as is the case for all focal-plane coronagraphs with small inner working angles. {High-frequency pointing jitter due to AO correction averages out to some level on an observing sequence of a few hours. Nevertheless, \rev{low-frequency drifts (typical frequency below $\sim$0.1~Hz)} in the centering of the star image onto the vortex phase mask may be caused by, for instance, noncommon path aberrations between the science channel and the AO wavefront sensing channel. These low-frequency drifts need to be controlled to a high accuracy to reach high contrast performance after ADI postprocessing.} Given the challenging science goal of the NEAR campaign, our goal in terms of pointing control was to reach a centering accuracy better than $\lambda/30D$ ($\sim$10~mas) over a 6-h observing sequence.


We adapted for NEAR observations the QACITS estimator\cite{Huby2015}{. This method is not wavelength-dependent in principle, and was already} validated on sky for the operation of vortex coronagraphs at shorter IR wavelengths (3--5~$\mu$m) on Keck/NIRC2\cite{Huby2017}. {However, its implementation for mid-IR observations is not straightforward nor guaranteed and requires dedicated simulations to verify its feasibility (higher background, see below) as well as several adjustments to its workflow, developed for shorter IR wavelengths (see Secs.~\ref{sec:qacitssimuls} and \ref{sec:qacitsnearimplement}).}

\begin{figure*}[t]
\centering
\includegraphics[width=.8\textwidth]{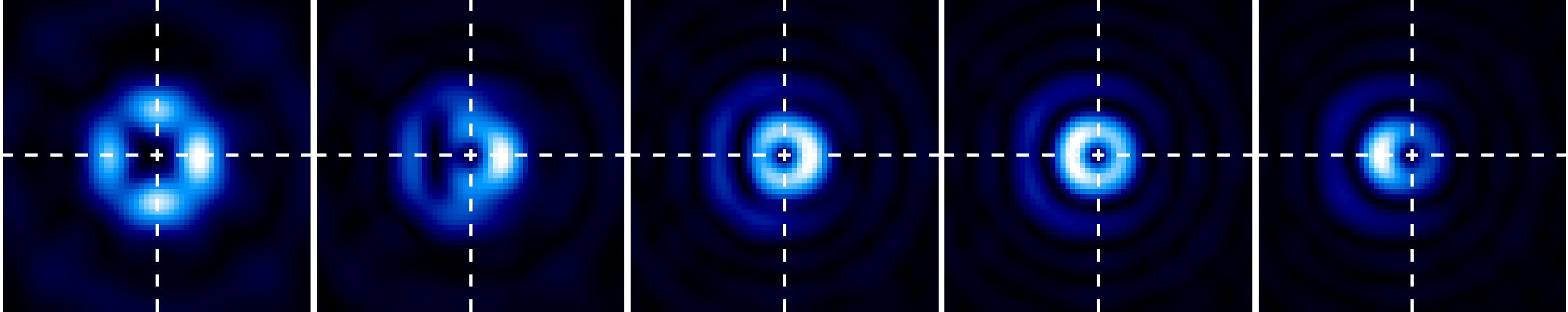}
\caption{Simulated VISIR images obtained with our apodized vortex coronagraph for different offsets applied to the star image toward the left: 0.01, 0.05, 0.10, 0.20, and 0.40~$\lambda/D$. The scale is square root for all images but the upper intensity cut is not the same (fraction of the peak intensity).}
\label{fig:offset_series}
\end{figure*}

The principle of QACITS is to use the coronagraphic images recorded on the science detector to estimate the offset of the star image with respect to the coronagraph. Figure~\ref{fig:offset_series} shows a series of simulated coronagraphic images for the NEAR configuration (VLT telescope pupil, apodized Lyot stop) with an increasing offset of the star toward the left. The dark central part of the coronagraphic pattern close to a perfect centering is due to the assumption of a perfect AGPM, with no intrinsic stellar leakage.

{The QACITS algorithm} estimates the centering error based on the measurement of the differential intensity between the four quadrants of the image. If only one direction is considered, the differential intensity corresponds to the flux difference between the two halves of the image.

Because of the central obscuration of the telescope pupil, there is no bijection between the coronagraphic differential intensity and the offset when looking at the innermost part of the coronagraphic pattern (Fig.~\ref{fig:offset_series}, first ring), which is the brightest part. In practice, to feed our QACITS estimator, we select and integrate only the outer part of the coronagraphic pattern (second ring), for which the intensity variation is a linear function of the offset of the star for small offsets ($\lesssim$0.5$\lambda/D$). \rev{The parameter in the QACITS model is a conversion factor from differential intensities to pointing offsets, which is estimated as the slope of a linear trend fit to measured differential intensities as a function of known pointing offsets.} For telescopes with large central obscurations such as Keck ($\sim$30\%), the intensity in the outer part of the coronagraphic PSF is adequate for the QACITS operation. In the case of the VLT (central obscuration $\sim$14\%), the second ring is significantly fainter, which degrades the sensitivity of QACITS to stellar offsets and leads to longer integration times for robust QACITS measurements.

In the case of mid-IR observations such as for NEAR, another potential issue for the QACITS operation arises from the significantly higher thermal background with respect to shorter IR wavelengths. We could not use on-sky VISIR data taken with the vortex coronagraph before the NEAR upgrade to validate the QACITS model, because the {conversion factor from differential intensity to offset strongly depends} on the shape of the telescope pupil and of the Lyot stop of the coronagraph, which were both different for the NEAR campaign on UT4. Also, on-sky data are not really suitable for such validation tests because we do not know the true value of the pointing offsets. Therefore, we relied on simulated data, where we introduced known pointing offsets, as described in the next section.

    \subsection{Expected performance}
    \label{sec:qacitssimuls}

In order to optimize the QACITS frame rate as a function of the S/N of the measurements, and to evaluate its expected performance for NEAR observations, we simulated NEAR data with the Python package HEEPS (\rev{high-contrast end-to-end performance simulator}, \url{https://github.com/vortex-exoplanet/HEEPS}). For all these simulations, we used the Lyot stop with the {least} aggressive design. {The S/N was measured as the ratio of the median signal in the annulus considered by QACITS and of the standard deviation in an annulus covering pixels at larger separation, representative of the background noise.}

\begin{figure*}[t]
\centering
\includegraphics[width=.9\textwidth, trim = 0mm 0mm 8mm 8mm, clip]{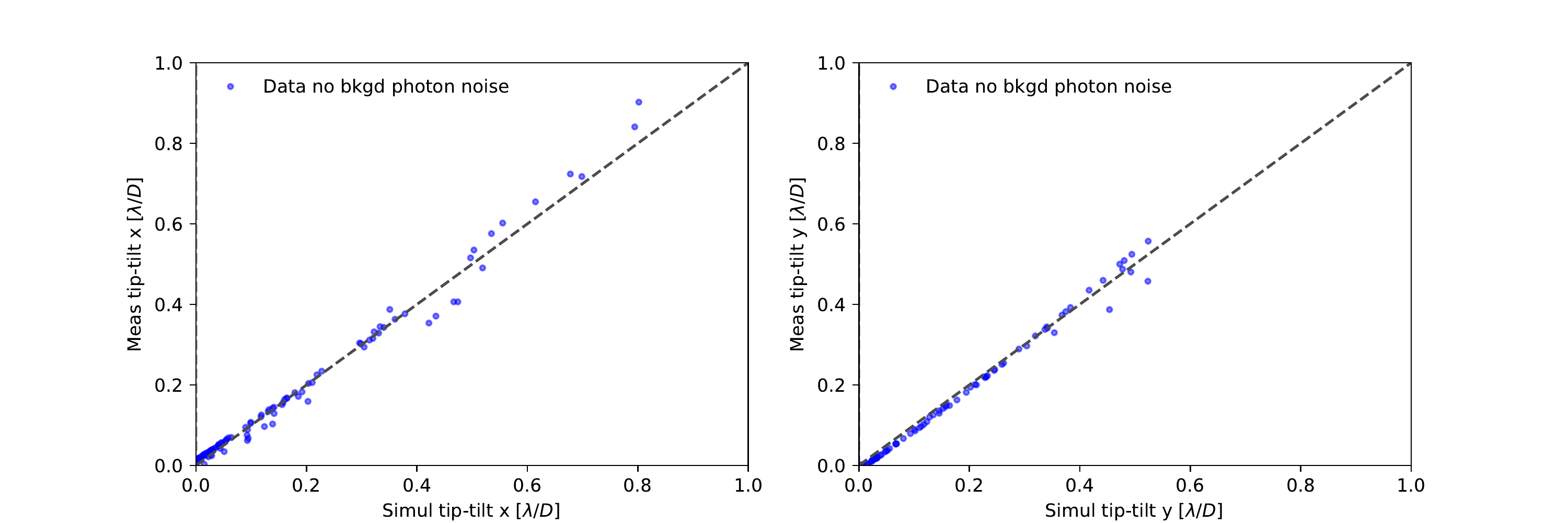}
\caption{{Comparison of the simulated and estimated offsets by QACITS along $x$ (\textit{left}) and along $y$ (\textit{right}) for the tests without photon noise associated with the thermal background.}}
\label{fig:qacitstestsretrieval}
\end{figure*}

First, we verified the correctness of the optimized \rev{conversion factor from differential intensity to pointing offset in the QACITS model} from simulated NEAR {images} by analyzing the fidelity of the offset estimation. For this, we simulated NEAR images for various known offsets without introducing photon noise associated with the thermal background. The comparison of the simulated and estimated offsets is shown in Fig.~\ref{fig:qacitstestsretrieval}. {We note that QACITS is able to retrieve the simulated offsets well. The statistics (mean and standard deviation) of the measured differences along the $x$ and $y$ directions gives $0.006 \pm 0.033 \, \lambda/D$ and $-0.004 \pm 0.047 \, \lambda/D$, respectively}. This test validates the values of the QACITS \rev{conversion factor}. {The measured dispersion for a given offset amplitude is due to the fact that the QACITS \rev{conversion factor was} estimated from simulated data with offsets applied in one direction only, whereas in our fidelity test we randomly drew the amplitude and the orientation of the offsets and the brightness distribution in the AGPM pattern slightly changes with the orientation of the offset.}

Then, we evaluated the requirements on the S/N in the region used by QACITS and on the integration times for robust QACITS measurements of the stellar offsets. For this, photon noise associated with the thermal background estimated from on-sky VISIR data taken with the early version of the vortex coronagraph (program ID: 097.C-0705, target: Fomalhaut) was added to the simulated images. {We used the same data generated for the fidelity analysis except that we restricted the analysis to offsets with amplitudes smaller than 0.1\,$\lambda/D$, representative of closed-loop operations. Offsets are more difficult to measure in this regime, because they are associated with a low flux in the outer coronagraphic ring used by QACITS. As a result, longer integration times are expected to be required to achieve a sufficient S/N on the offset measurement.}

\begin{figure*}[t]
\centering
\includegraphics[width=.9\textwidth, trim = 0mm 0mm 8mm 8mm, clip]{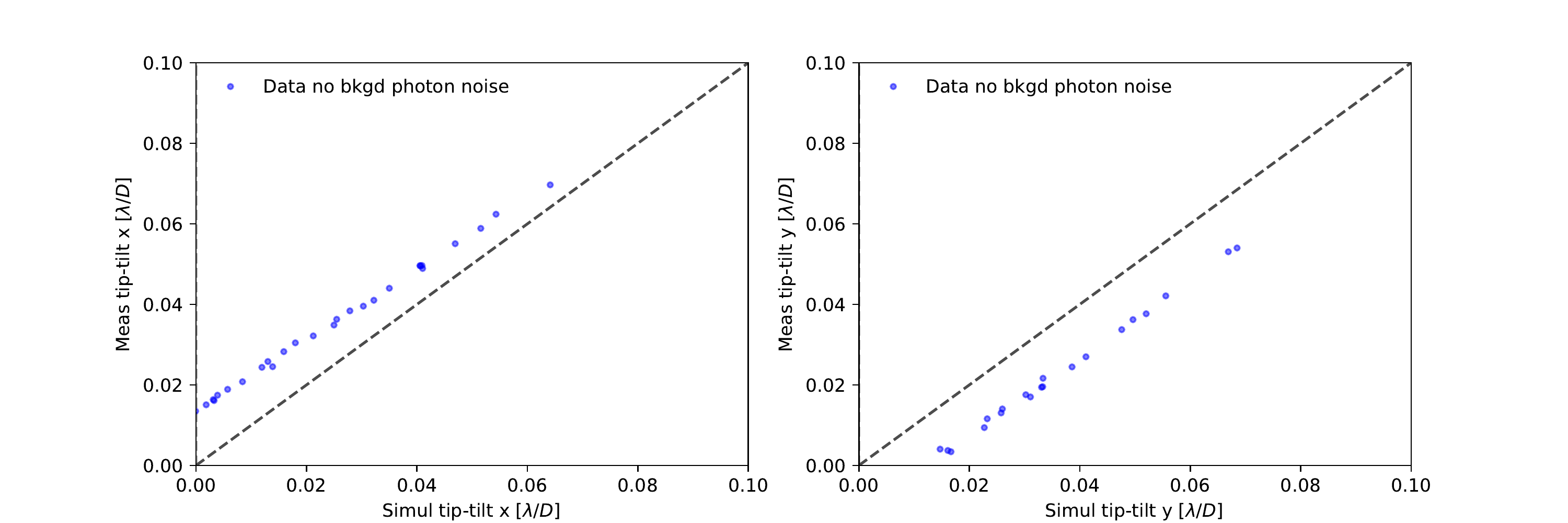}
\caption{{Comparison of the simulated and estimated offsets by QACITS along $x$ (\textit{left}) and along $y$ (\textit{right}) for the test without photon noise associated with the thermal background (zoomed version of Fig.~\ref{fig:qacitstestsretrieval}), focused on offsets smaller than 0.1\,$\lambda/D$ (see text).}}
\label{fig:qacitstestsretrievalnoiseref}
\end{figure*}

\begin{figure*}[t]
\centering
\includegraphics[width=.9\textwidth, trim = 0mm 0mm 8mm 8mm, clip]{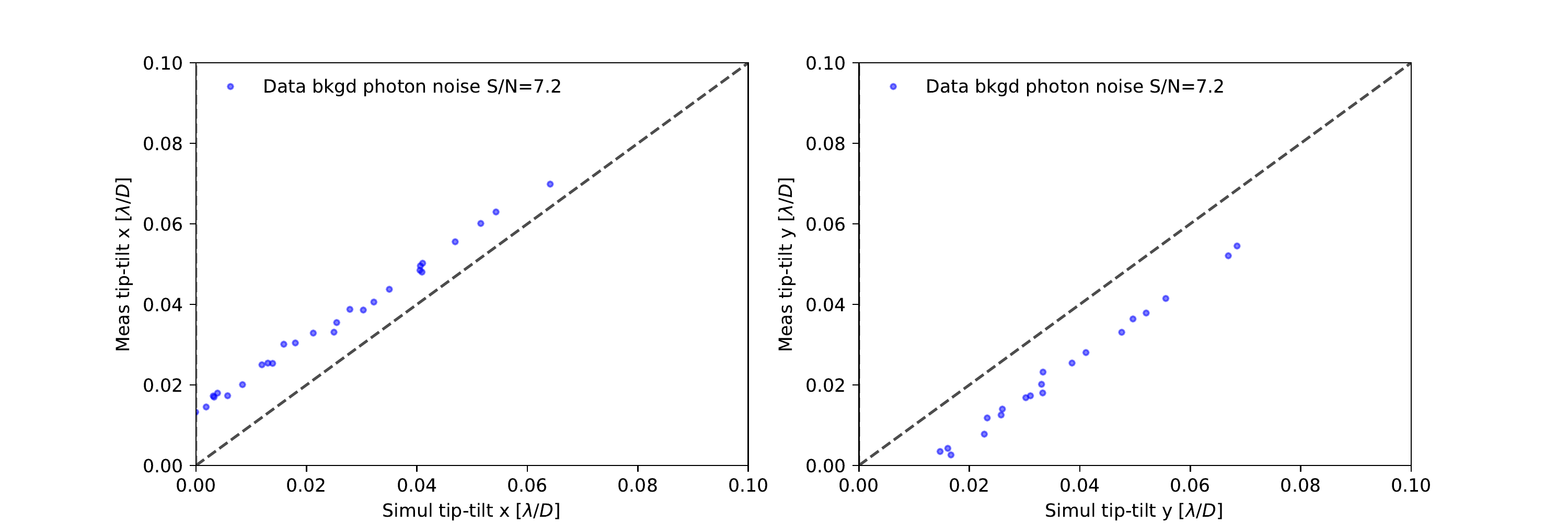}
\caption{{Same as Fig.~\ref{fig:qacitstestsretrievalnoiseref} but for the test with background photon noise.}}
\label{fig:qacitstestsretrievalnoise}
\end{figure*}

{We show in Figs.~\ref{fig:qacitstestsretrievalnoiseref} and \ref{fig:qacitstestsretrievalnoise} the comparison of the simulated and estimated offsets in the case without photon noise
associated with the thermal background and in the case with background photon noise for a S/N of $\sim$7, respectively. This S/N is achieved in $\sim$90~ms. The statistics (mean and standard deviation) of the measured differences along the $x$ and $y$ directions gives $0.0132 \pm 0.0036 \, \lambda/D$ and $-0.0117 \pm 0.0020 \, \lambda/D$, respectively. For comparison, the statistics for the noiseless measurements is $0.0133 \pm 0.0036 \, \lambda/D$ and $-0.0116 \pm 0.0018 \, \lambda/D$, respectively. Thus, QACITS is able to retrieve the simulated offsets even in presence of background noise, at a high cadence ($\sim$10\,Hz). This cadence is expected to be largely sufficient for the control of pointing drifts in NEAR observations, which benefit from the good stability of the AGPM position inside the cryostat and from the use of AO to stabilize the position of the star image behind the AGPM. The results from these tests also suggest that the star image centering control based on QACITS may be biased at the level of $\sim$0.01~$\lambda/D$ in each direction. We checked with additional noiseless simulations that the bias is present for a perfect centering of the star image onto the vortex phase mask, and that it is caused by the M3 shadow in the telescope pupil, which breaks its symmetry (Fig.~\ref{fig:apodstops}). In conclusion, our simulations suggest that photon noise associated with the thermal background will not be an issue for QACITS operations on VISIR-NEAR.}

\begin{figure}[t]
\centering
\includegraphics[width=.99\textwidth, trim = 22mm 0mm 23mm 0mm, clip]{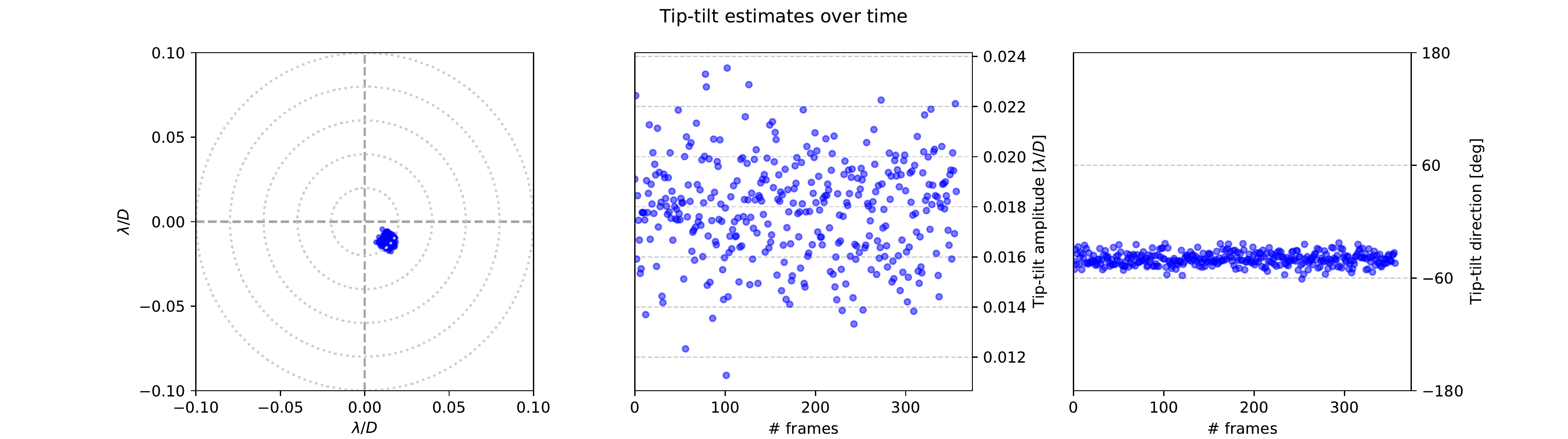}
\caption{{Offsets measured by QACITS due to simulated AO residual aberrations in the $x$-$y$ plane, in amplitude as a function of the frame ID, and in direction as a function of the frame ID (see text).}}
\label{fig:qacits_testao}
\end{figure}

Finally, we evaluated the impact of nonperfect AO correction on the QACITS performance. QACITS is not intended to estimate fast pointing jitter due to atmospheric turbulence, so the accuracy down to which we can control the star image centering might be limited by the ability of the AO to keep the star image behind the AGPM at frequencies above the frequency of the QACITS-based loop for offset correction (typically 0.1~Hz). For this analysis, we simulated AO phase screens with a configuration similar to the AO module used for NEAR using the COMPASS package\cite{Gratadour2014} (\url{https://github.com/ANR-COMPASS}). We used these phase screens as an input for the HEEPS simulator, and generated simulated NEAR images without additional pointing offsets. {We also added photon noise associated with the thermal background as described in the previous paragraph.} These images were then analyzed with QACITS to estimate the pointing errors. Figure~\ref{fig:qacits_testao} shows the estimated pointing errors for an image binning corresponding to a time resolution of 3\,s. We note that the {largest amplitudes measured by QACITS are $\sim$0.024$\lambda/D$. The statistics {(mean and standard deviation)} of the measured {differences between the measured and simulated offsets along the $x$ and $y$ directions} gives $0.014 \pm 0.002 \, \lambda/D$ in $x$ and $-0.011 \pm 0.002 \, \lambda/D$ in $y$. We recover the bias of $\sim0.01 \, \lambda/D$ in each direction, due to the M3 shadow in the VLT pupil. This analysis suggests that the quality of the correction of the NEAR AO should not affect the QACITS performance.} For the NEAR campaign, QACITS was used to estimate stellar offsets every 30\,s. {In order to account for the bias due to the M3 shadow in the on-sky observations, we added a set point parameter to the original QACITS model (Sec.~\ref{sec:onskyqacitsclosedloop}), which was optimized manually during on-sky operations by minimizing the intensity of the stellar leakage.}

    \subsection{Implementation on NEAR}
    \label{sec:qacitsnearimplement}

We describe here the practical implementation of QACITS for NEAR. {With respect to the procedure developed for $L^{\prime}$-band AGPM observations\cite{Huby2017}, several modifications were made to account for the specificities of mid-IR observations (high, time-variable thermal background) and of the observing strategy of NEAR (chopping between the two stars of $\alpha$ Cen, using pupil tracking).} The workflow is the following:
\begin{enumerate}
    \item At the beginning of the night, the position of the AGPM center is updated in the QACITS model by measuring its on-sky position as seen by the NEAR detector (see Sec.~\ref{sec:vortexcenter}).
    \item A series of background images is recorded in chopping mode. They are chop-subtracted and median-combined to estimate the intensity and structure of the background residuals after chopping (see Sec.~\ref{sec:bkgdcalib}).
    \item The science images on $\alpha$~Cen are recorded in chopping mode, chop-subtracted, and median-combined over a few seconds. The chopped background image is removed from that science image (after scaling if needed) to provide the input image to the QACITS estimator.
    \item The QACITS estimator extracts from the input image the aperture photometry of the noncoronagraphic PSFs of the A and B stars, and subtracts them to get the photometric normalization for the coronagraphic data. {An advantage of mid-IR chopped images is that nonsaturated measurements of the stellar photometry is obtained in real time during the whole sequence, while dedicated photometric measurements are required to feed QACITS at shorter wavelengths\cite{Huby2017}.}
    \item QACITS then measures the differential intensities in the areas of the outer coronagraphic ring, and estimates the stellar offsets in the $x$ and $y$ directions from the normalized differential intensities.
    \item These stellar offsets are converted into correction offsets, and sent to the field selector of the telescope.
\end{enumerate} 

In closed-loop mode, the last four steps are repeated for about an hour. Every hour, new background images are recorded (step 2{, see Sec.~\ref{sec:bkgdcalib}}), and the model for the background residuals is updated in order to capture the slow background variations.

\begin{figure}[t]
\centering
\includegraphics[width=.44\textwidth]{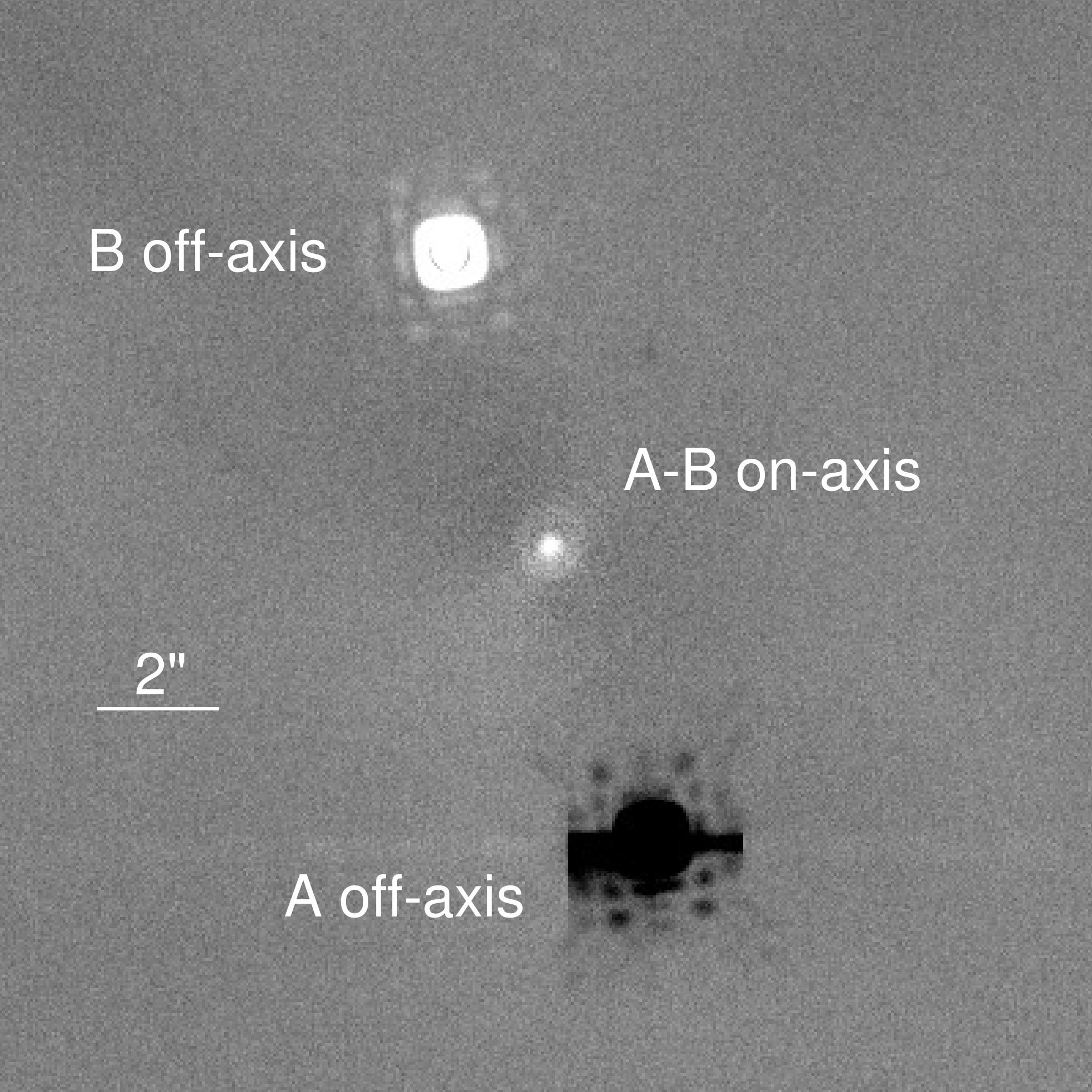}
\caption{Typical NEAR image obtained after chopping and median combination over a single data cube (useful integration time 24\,s). The chopping residuals were also subtracted. The horizontal stripe seen on the off-axis PSF of $\alpha$ Cen A is due to a detector defect. See text for details.}
\label{fig:nearchoppedim}
\end{figure}

Figure~\ref{fig:nearchoppedim} shows a typical input image for QACITS, with the vortex coronagraph centered on the optical axis in the middle of the image, and the off-axis PSFs of $\alpha$~Cen A (negative) and of $\alpha$~Cen B (positive) on either side. Due to chopping, the coronagraphic PSF is the difference between the coronagraphic PSFs of $\alpha$ Cen A and B (the flux ratio between the two stars is $\sim$40\%). Due to the apodized Lyot stop (Sec.~\ref{sec:apodlyotstops}), the diffraction rings and diffraction patterns due to the telescope spiders around the off-axis PSFs in the range from 2.5$''$ to 7.5$''$ are highly attenuated.

We describe below in further details two specific steps of the QACITS workflow outlined above, namely the vortex center position measurement and the background subtraction, {which are specific to this instrument and wavelength regime}.

        \subsubsection{Vortex position measurement}
        \label{sec:vortexcenter}

The accurate knowledge of the position of the AGPM center with respect to the science detector is critical for QACITS operations, as it defines the reference position for the QACITS measurements. Fortunately, the position of the AGPM center can be readily identified from a simple flat field (sky background) measurement. In the case of VISIR, where there is no cold pupil stop upstream of the AGPM, thermal emission located outside the entrance pupil is partly diffracted back into the telescope pupil by the vortex phase mask, leading to the ``vortex center glow'' effect\cite{Absil2016}. The structure of this glow, which is about twice as bright as the local background emission in the case of VISIR, can be approximated by a superposition of two 2D Gaussian patterns, associated respectively with thermal emission coming from outside and inside (i.e., central obscuration) the geometric image of the VLT primary mirror. An automatic procedure based on 2D Gaussian fitting was developed to measure the position of the glow of the AGPM at the beginning of each NEAR campaign night. {For QACITS to work without bias, the position of the AGPM center has to be stable during the whole observation. Instabilities could be caused by differential flexures or thermal drifts inside the instrument. Fortunately for NEAR, the} AGPM position {as seen from the detector} was shown to be stable within $\lesssim 0.5$~pix ($\lesssim 22$~mas) based on tracking tests performed with the telescope dome closed during the integration of NEAR, using a trajectory similar to the trajectory expected for $\alpha$ Cen. Measurements of the AGPM center position during the first nights of the NEAR campaign confirmed the subpixel stability of the AGPM center position.

        \subsubsection{Background subtraction}
        \label{sec:bkgdcalib}

During the first QACITS tests with on-sky data, we discovered that the offset estimates were affected by a bias of $\sim$0.2~$\lambda/D$, well beyond the NEAR requirements. This bias originated from background subtraction residuals in the images sent to QACITS. {To maximize the duty cycle of the NEAR campaign, it was indeed not possible to exploit the standard chopping-and-nodding scheme for mid-IR observations.} Instead, the background subtraction scheme for NEAR is based on pure chopping between the two $\alpha$ Cen stars, where each star is alternately centered on the AGPM.

\begin{figure}[t]
\centering
\includegraphics[width=.4\textwidth]{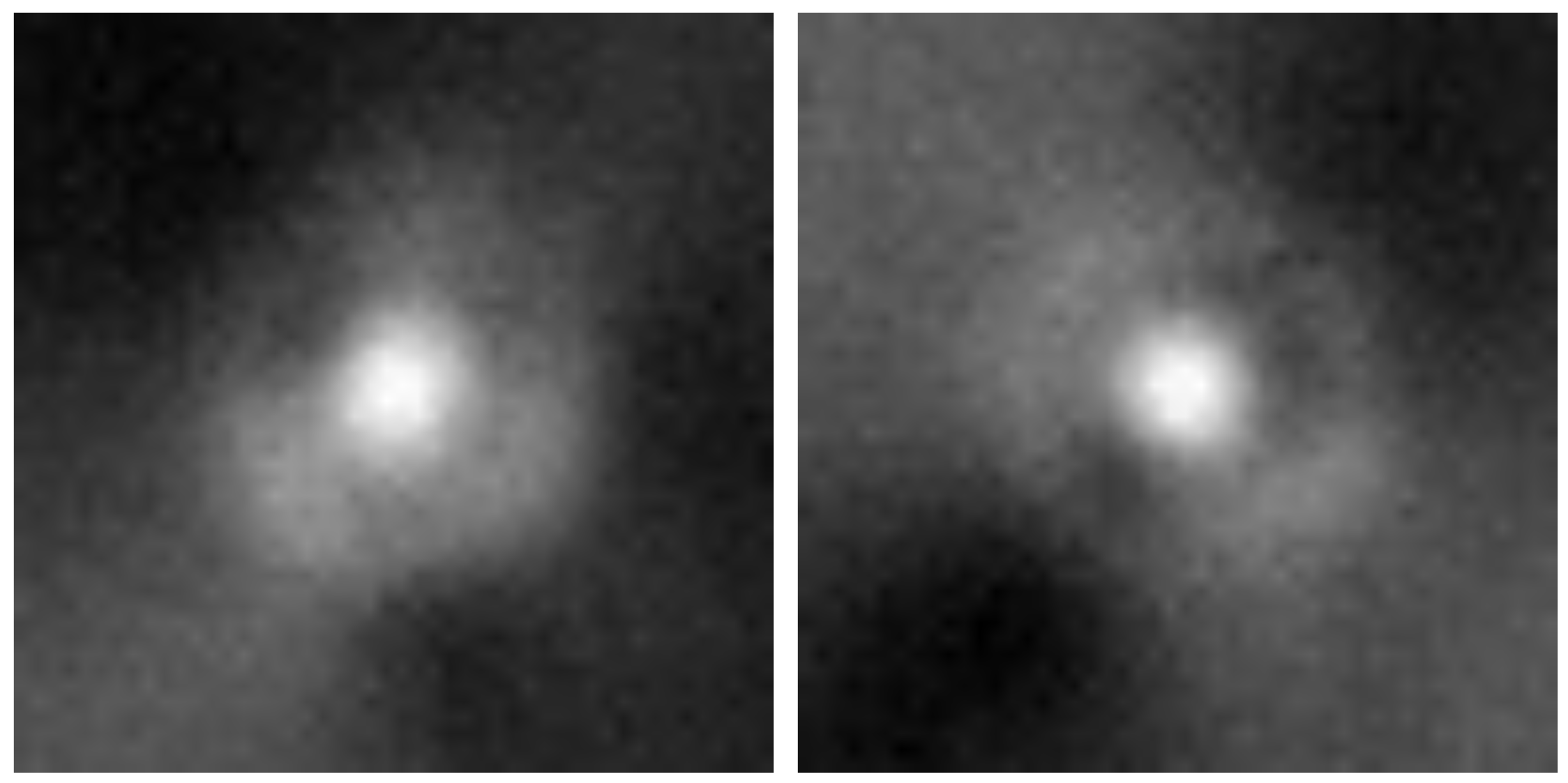}
\caption{NEAR chopped image zoomed around the coronagraph center, obtained at the beginning (\textit{left}) and at the end (\textit{right}) of a 6-h ADI sequence.}
\label{fig:chopbkgdres}
\end{figure}

In order to subtract the background residuals in the chopped images before they are sent to QACITS, we adjusted the observing strategy to acquire background calibration data in chopping mode on an empty sky region. However, when analyzing the data further, it appeared that the background residuals in the chopped images were not stable, but evolved with a typical timescale of an hour due to the use of the pupil-tracking mode. The background variation between the beginning and the end of a 6-h ADI sequence is shown in Fig.~\ref{fig:chopbkgdres}. Because of this feature, it was necessary to acquire background calibration data regularly (every $\sim$1h or so). {The lessons learned from this analysis are valuable for the implementation of QACITS for ELT/METIS observations, which will likely make use of a similar observing strategy for background subtraction.}

\section{On-sky results}
\label{sec:onskyresults}

The VISIR-NEAR commissioning spanned ten half nights on 2019 April 3--13 UT and two additional half nights on 2019 May 21--23 UT to complete the coronagraphic and QACITS tests. Before proceeding with the QACITS tests, the whole instrument (including the AO module) had to be tested and optimized. The performance of phase mask coronagraphs such as the AGPM is sensitive to the presence of low-order aberrations such as defocus and astigmatism. To mitigate the impact of these aberrations, they were carefully measured and calibrated in the first nights of the commissioning. 

First, the AO loop was optimized with respect to two parameters: the number of controlled Karhunen-Loeve modes and the feedback gain of the control loop. The maximum number of modes that can be controlled with the 1170 actuator deformable secondary mirror (DSM) is around 800. The required force to be applied to the DSM actuators, however, increases rapidly with the mode number, i.e., with the spatial frequency that is to be controlled\cite{riccardi06}. For NEAR, the DSM is not only used {to correct for turbulence and telescope tracking errors}, but also provides {the chopping throw of 5$''$ (angular separation between $\alpha$ Cen A and B at the time of the observations)}. To avoid actuator saturation, we, therefore, restricted the number of controlled modes to about 350. This still provided an N-band Strehl ratio {larger than 95\% typically}, and let us control all the low spatial frequencies up to angular separations of about 2$''$ (7~$\lambda/D$), in which we were scientifically interested. The feedback gain of the control loop was set to the nominal value of 50\% typically used by the AOF systems at the VLT.

Second, we had to ensure that the {noncommon path between the science channel and the wavefront sensing channel of NEAR} did not induce significant additional aberrations. The noncommon path of NEAR is quite simple, because the coronagraphic mask is the first optical element behind the dichroic used to split off the optical light and direct it to the AO wavefront sensor. The dichroic substrate is a 10-mm thick ZnSe window at 45 degrees in the f/13.4 beam of the VLT Cassegrain focus. {It introduces} mainly about 300~nm rms of astigmatism. Other modes contribute at the level of only a few tens of nm rms, which is negligible in the N band. Noncommon path aberrations in the wavefront sensing arm were calibrated by measuring a reference signal with an optical fiber inserted in the input focus. We also installed a ZELDA mask (\rev{Zernike sensor for extremely low-level differential aberrations})\cite{Ndiaye14} in the VISIR focal plane wheel, which measured and {confirmed the dominant astigmatism as expected}. Finally, we also measured defocus and astigmatism curves on-sky, i.e., we recorded the AO-corrected PSF sweeping through a range of defocus and astigmatism offsets to determine the values where the PSF is best. These values were again consistent with our expectations and used throughout the observing campaign.

During these tests, the Lyot stop with the {least} aggressive design was also selected for the campaign. This choice was motivated by the fact that the contrast performance is quickly limited by the instrument background for larger angular separations, and that achieving a higher throughput was more relevant at that point.

\subsection{Coronagraphic performance}
\label{sec:onskyagpmperf}

\begin{figure}[t]
\centering
\hspace{10mm}
\includegraphics[width=.69\textwidth, trim = 3mm 4mm 5mm 8mm, clip]{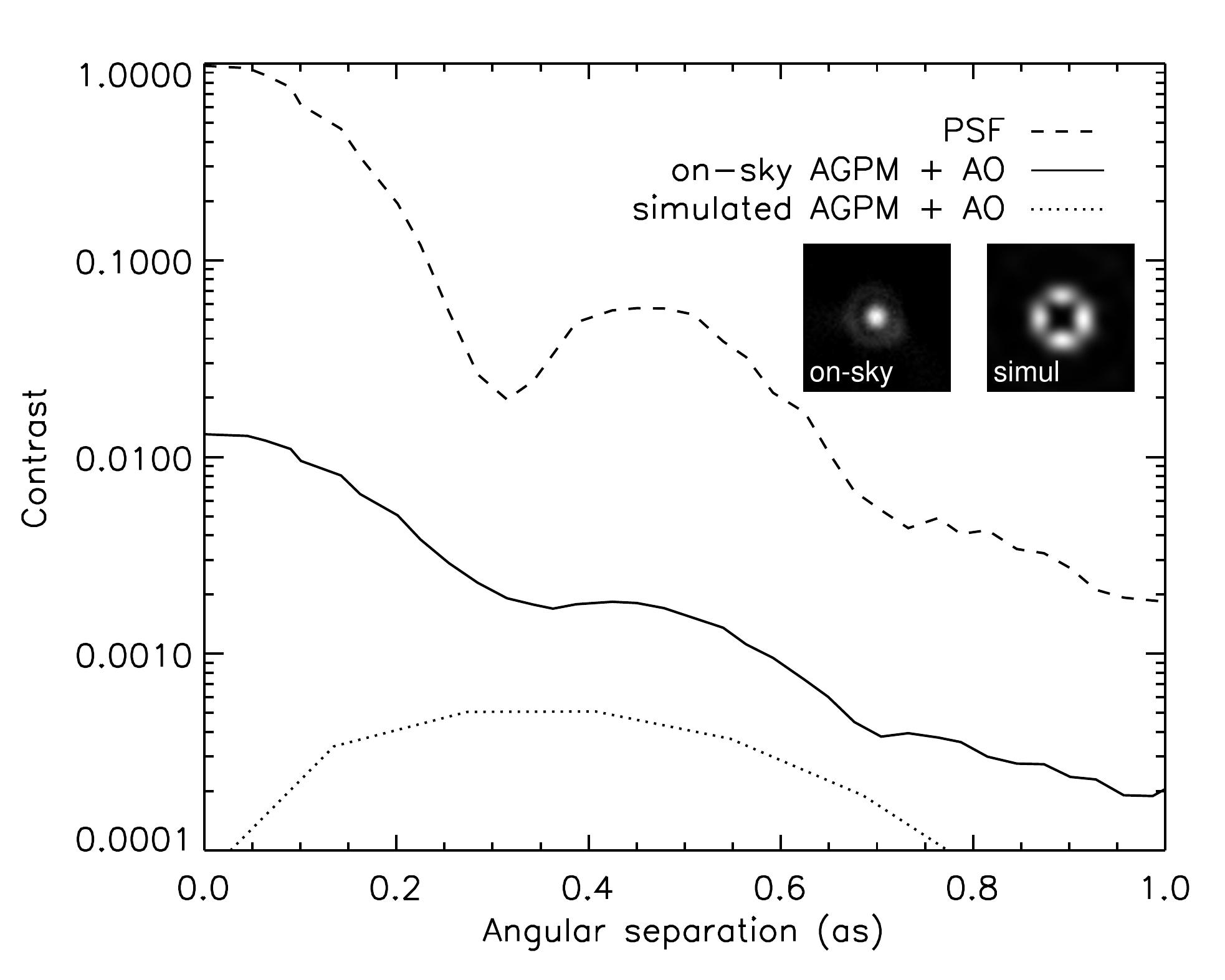}
\caption{{Radial profiles averaged azimuthally for an unsaturated noncoronagraphic NEAR PSF, NEAR AGPM data recorded with the QACITS-based loop closed, and a perfect AGPM close to perfect centering with simulated AO residuals (see text and inset images).}}
\label{fig:rejection_perfs}
\end{figure}

We measured on-sky the rejection performance of the three AGPMs installed in VISIR: (i) AGPM-N4, the original mask installed in VISIR in 2012, (ii) AGPM-N3, a device manufactured in 2012 but not installed in VISIR at that time, and (iii) AGPM-BT3, the mask manufactured specifically for the NEAR campaign that showed the best performance {among the newly manufactured masks during} our laboratory tests (Sec.~\ref{sec:agpm_masks}). None of the three AGPMs reached the expected on-sky performance derived from our simulations of residual turbulence after AO correction representative of the VLT AOF (Sec.~\ref{sec:qacitssimuls}), using a perfect AGPM (see dotted curve in Fig.~\ref{fig:rejection_perfs}). Instead, they provided a maximum rejection rate of about 100:1, integrated over the NEAR filter, with AGPM-N4 turning out to show the best performance in our on-sky tests {(see solid curve in Fig.~\ref{fig:rejection_perfs})}.

The measured radial profile for the coronagraphic PSF strongly deviates from our \rev{simulated} predictions, especially within $1\lambda/D$, where the radial profile shows a peak instead of a dark hole. (Note that separations smaller than $1\lambda/D$ present little scientific interest because of the intrinsic AGPM extinction.) {This was the case for all three AGPMs.} \rev{We note that our \rev{simulated} performance does not account for the intrinsic rejection performance of the AGPM (assumption of a perfect AGPM). We verified through additional simulations that low-order aberrations (defocus, astigmatism) cannot reproduce a bright central peak. A large astigmatism ($>$400~nm rms) could produce a bright central peak, but it would have been measured with the ZELDA mask (Sec.~\ref{sec:onskyresults}). The measured AGPM profile} is the telltale sign of a nonperfect AGPM. The reason why it was not possible to reach a rejection rate of 400:1, as measured for AGPM-N3 in the lab, is not entirely clear, although two main explanations can be proposed: (i) AGPM-N3 was only tested around 10.5~$\mu$m on our test bench, and it could turn out that its performance strongly degrades at longer wavelengths; (ii) as described in Ref.~\citenum{Kasper2019}, a leak in the NEAR cryostat has led to nitrogen ice entering the grooves of the AGPMs, which may have significantly degraded the AGPM performance from the very beginning of our commissioning campaign. A significant gain (up to a factor $\sim 4$) in terms of raw contrast could potentially be reached by fixing one or both of these issues in the future. Fixing issue (i) would require a more extensive testing facility, which is currently being developed at CEA Saclay.

\subsection{Validation of the QACITS model}
\label{sec:onskyqacitsvalid}

\begin{figure*}[t]
\centering
\includegraphics[width=.99\textwidth, trim = 0mm 0mm 0mm 0mm, clip]{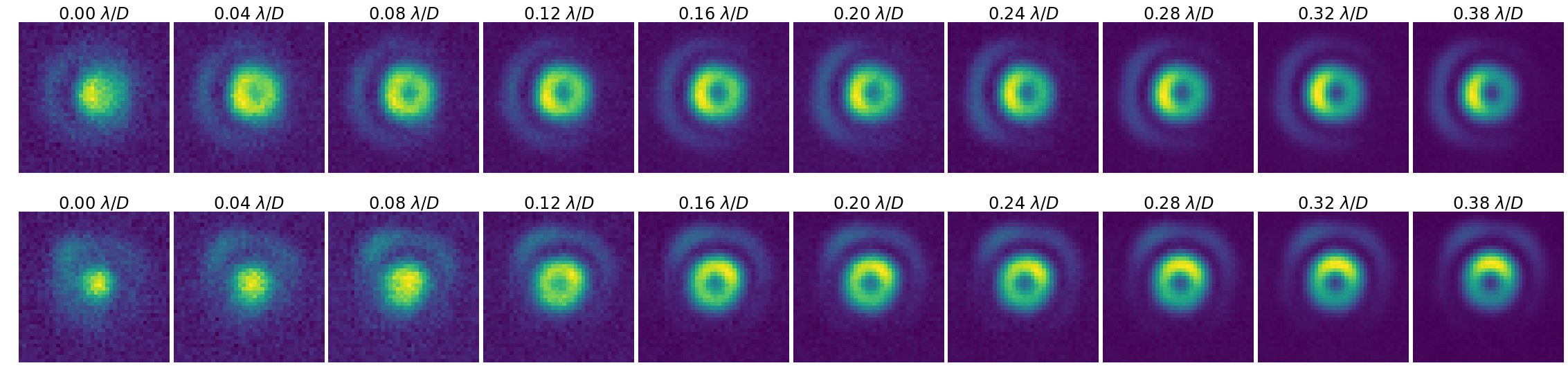}
\caption{{Measured coronagraphic} PSF for various offsets applied in two orthogonal directions: toward the left (top row images) and upward (bottom row images). {The observed source is $\alpha$ Cen}. First image on the left corresponds to the best centered position, and last image on the right to the largest offset applied ($0.38\lambda/D$). {The images were rescaled in intensity differently.}}
\label{fig:qacits_ramps}
\end{figure*}

To validate the QACITS model on sky, we applied with the field selector of the telescope a series of known offsets in two orthogonal directions. Figure~\ref{fig:qacits_ramps} shows a subsample of the series of recorded coronagraphic images. We note that the coronagraphic PSF has a peculiar shape in the on-sky images for small pointing errors (first images from the left in Fig.~\ref{fig:qacits_ramps}) with respect to our simulations (Fig.~\ref{fig:offset_series}), and to what is seen in other imaging instruments equipped with vortex coronagraphs, operating at shorter IR wavelengths (3--5~$\mu$m)\cite{Absil2013, Defrere2014, Serabyn2017}. The coronagraphic PSF does not look like the standard ``donut'' shape, but rather like a scaled-down version of the noncoronagraphic PSF. This pattern is in fact expected from theory\cite{Mawet2005}, in the case where the coronagraphic performance is limited by the intrinsic performance of the vortex phase mask, rather than by the residual wavefront aberrations after AO correction or by diffraction effects due to the central obscuration. This is a confirmation of the excellent performance of the AO correction measured during the NEAR commissioning (Strehl ratios $>$95\%), but also of the subpar performance of the AGPMs. 

The actual structure of the coronagraphic PSF has practical consequences on the QACITS estimator, because the contribution of the intrinsic leakage from the AGPM moves with the star image position. We also note the absence of an inversion with respect to the coronagraph center for the location of the intensity peak in the inner coronagraphic ring, as seen in the simulated NEAR data of Fig.~\ref{fig:offset_series}. The \rev{conversion factor from differential intensity to pointing offset of the QACITS model was,} therefore, reoptimized using the on-sky test data \rev{by fitting a linear trend to the measured differential intensities as a function of the known pointing offsets}. The linear behavior of the QACITS estimator based on the second ring of the coronagraphic PSF made this reoptimization quite straightforward. The situation would have been much more complicated if the QACITS estimator for NEAR was based on the first ring in the coronagraphic PSF.

\begin{figure*}[t]
\centering
\includegraphics[height=.47\textwidth, trim = 2mm 0mm 3mm 0mm, clip]{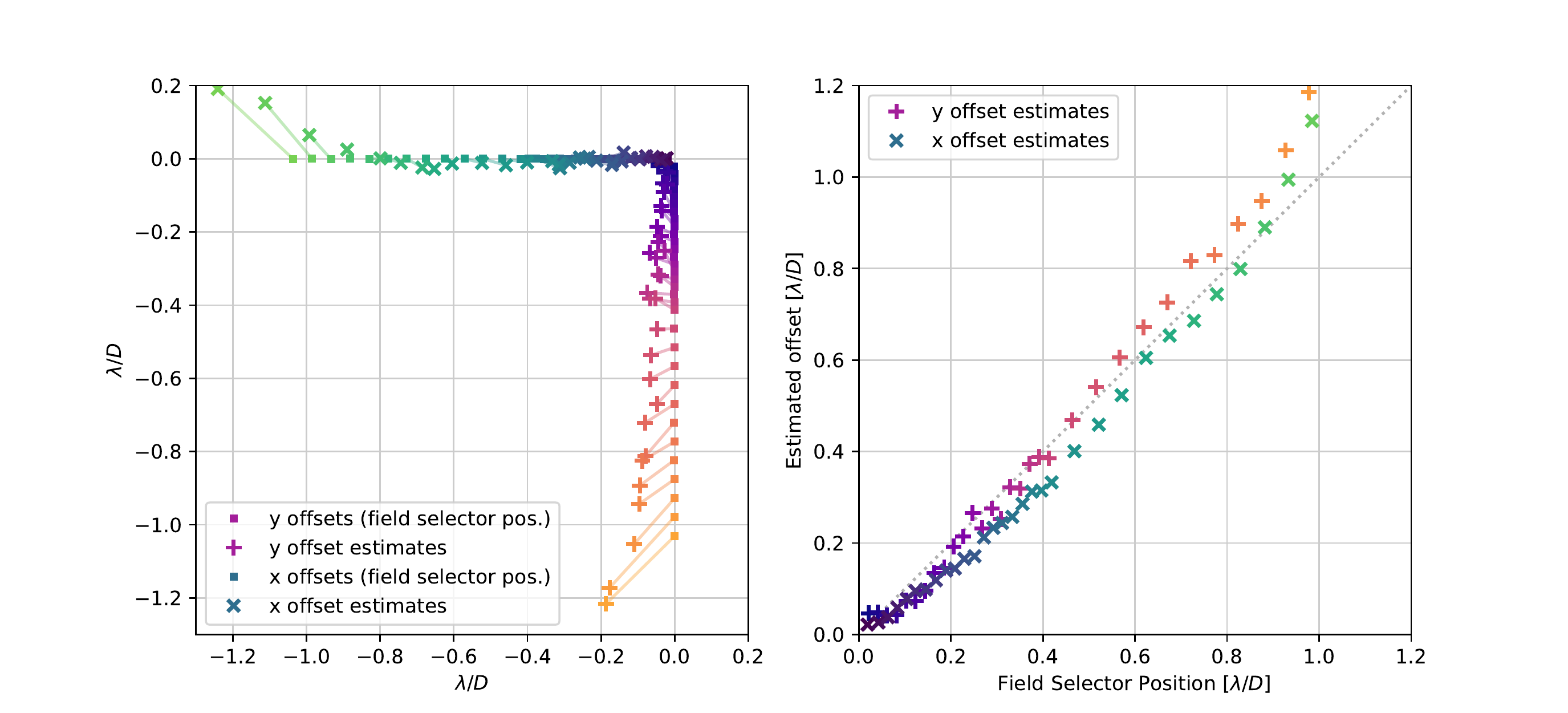}
\caption{Comparison of the offsets applied with the field selector of the VLT and those estimated by QACITS in the $x$-$y$ plane (\textit{left}) and measured offset as a function of the applied offset (\textit{right}).}
\label{fig:qacits_perfs}
\end{figure*}

Figure~\ref{fig:qacits_perfs} compares the applied and measured offsets, after reoptimization of the \rev{conversion factor from differential intensity to offset of the QACITS model}. We note the good agreement between the offsets up to values of $\sim$0.9$\lambda/D$ in $x$ and $\sim$0.5$\lambda/D$ in $y$. QACITS overestimates the offsets for larger values{, because the linear model that we assumed is not a good approximation of the real behavior anymore}. This is not an issue as we expect QACITS to typically work in a regime of small offsets ($\lesssim 0.2 \lambda/D$), and because this bias does not prevent from closing the loop.


\subsection{QACITS closed-loop performance}
\label{sec:onskyqacitsclosedloop}

In order to recenter the position of the star image behind the AGPM in real time based on the QACITS offset estimates, we implemented a feedback loop driving the position of the field selector of the telescope. With respect to the original QACITS model, we added a set point parameter to account for a small bias in the QACITS measurements when the star image is close to a perfect centering (Sec.~\ref{sec:qacitssimuls}). {After adjusting it manually by minimizing the stellar leakage at the beginning of each night, the set point was used for the entire night, with only small adjustments.} In addition, we added a gain factor in the conversion of the QACITS offset estimates into the offsets applied to the field selector. The controller gain was optimized as follows. A series of step-like perturbation signals were used to excite the system while operated in closed loop with a number of different candidate controllers. These data were processed to estimate a simple model representing the dynamic response of the system, i.e., the mapping from field selector mirror offsets to the output of the QACITS estimator. Such a model was then used to derive a simple dynamic model of the response of the system and in turn to select appropriate controller parameters achieving a satisfactory trade-off between performance and robustness.

\begin{figure}[t]
\centering
\includegraphics[width=.42\textwidth, trim = 2mm 2mm 2mm 2mm, clip]{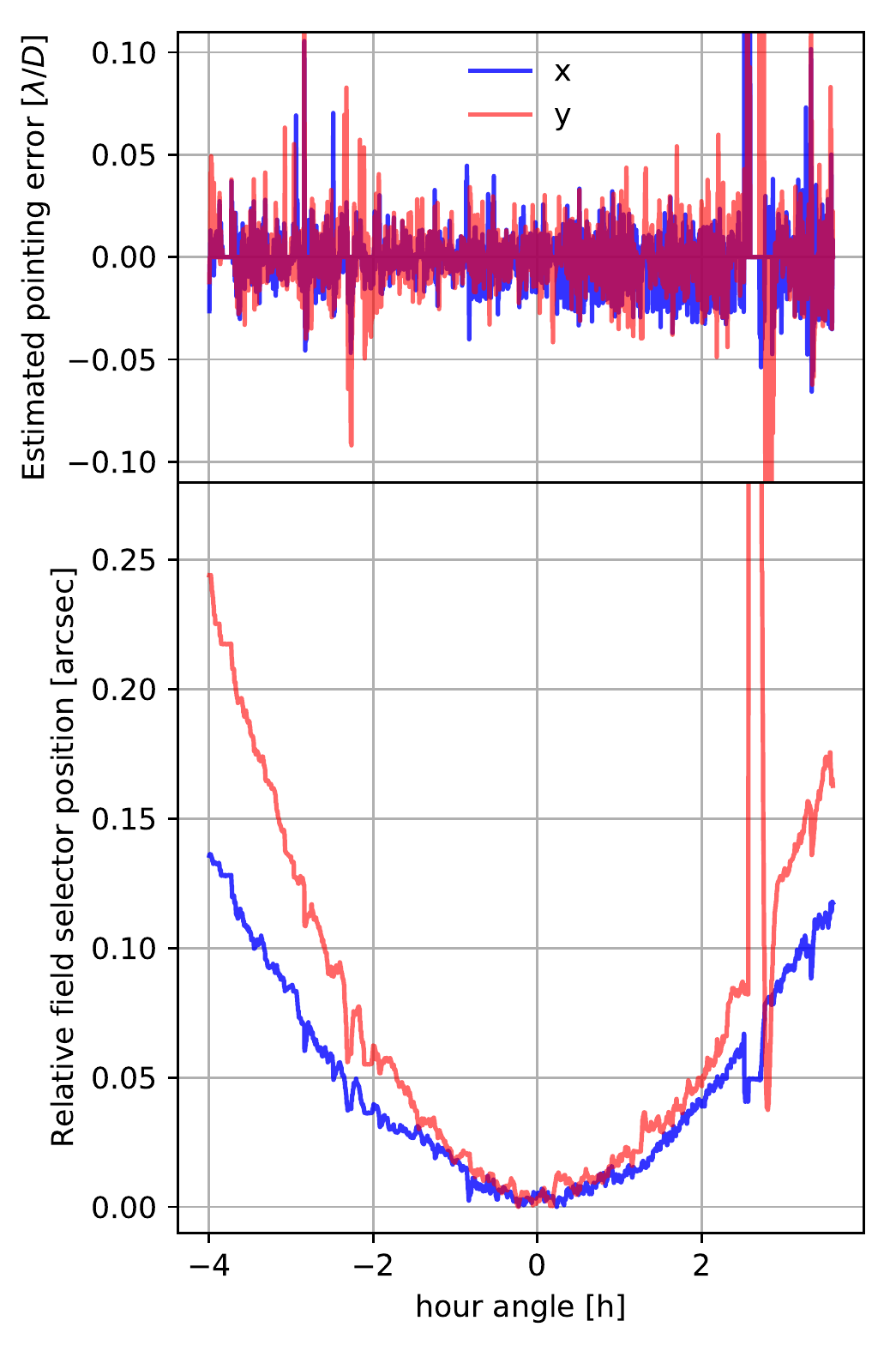}
\caption{{Time evolution of the estimated pointing error (\textit{top}) and of the relative position of the field selector (\textit{bottom}) measured on $\alpha$ Cen during the NEAR campaign night of 2019 May 25 UT}. An AO interruption occurred at hour angle $\sim$2.5 to 3~h.}
\label{fig:field_selector}
\end{figure}

Figure~\ref{fig:field_selector} shows the results obtained in closed loop for a sequence recorded during one of the best nights of the NEAR campaign, in terms of atmospheric turbulence and background emission level. The statistics of the estimated pointing errors (top panel) show that {the QACITS-based loop for offset correction} is able to control the AGPM centering over four hours around meridian passage with a stability of $0.015 \lambda/D$ rms. {For comparison, the statistics over 2-h intervals are: $0.016 \lambda/D$ rms along $x$ and $0.017 \lambda/D$ rms along $y$ for the hour angle range [0,2), $0.012$ and $0.013 \lambda/D$ rms for the range [-2,0), and $0.018$ and $0.027 \lambda/D$ rms for the hour angle range [-4,-2).} We also analyzed the relative position of the field selector over the night (bottom panel) and found {parabolic shapes} typical of differential atmospheric refraction. This behavior was expected given that the wavefront sensing is performed in the red part of the $I$ band (0.8--0.95~$\mu$m) with a sensor module installed inside VISIR, whereas the science observations are performed in the range 10--12.5~$\mu$m, without any atmospheric dispersion compensation.

As expected, the timescale of atmospheric refraction variation depends on the time to meridian passage (elevation above the horizon). It is slow when observing close to meridian passage (the elevation above the horizon is maximal) and gets faster when the target is farther from meridian passage (the elevation above the horizon gets smaller). Over four hours before the meridian passage, the position of the field selector moved by $\sim 1 \lambda/D$, whereas over two hours before/after meridian passage, it moved by $\sim 0.25 \lambda/D$. This behavior has important consequences on the offset correction accuracy, as the correction loop is always lagging behind to some level \rev{and that we chose to use a constant correction frequency for the implementation for the NEAR campaign}. It implies that the offset correction accuracy changes as a function of the elevation, with improving performance when the elevation above the horizon is maximal. This is confirmed by the evolution of the estimated pointing error, where the peak-to-peak variations are smaller when observing closer to meridian passage. However, the increase in the peak-to-peak variations stays rather small over the whole night, meaning that the closed-loop bandwidth is adequate to correct for differential atmospheric refraction. \rev{This is not an issue due to QACITS because it estimates the offsets. One way to compensate for the degradation of the offset correction accuracy would be to use a higher frequency for the correction loop.} 

In principle, observing close to meridian passage to maximize the field rotation for ADI is the optimal approach for high-contrast imaging observations. For NEAR, we decided to relax this constraint so that we could reduce the number of consecutive nights required to complete the campaign and thus minimize the smearing of the signal of putative planets due to orbital motion when combining all the data in the high-contrast analysis.

\section{Conclusions}

We presented the on-sky demonstration of precise centering control of the star image onto a vortex coronagraph {for mid-IR observations}, using the VISIR-NEAR instrument. {Although the QACITS method that we used is not wavelength-dependent, its implementation for mid-IR observations was not straightforward nor guaranteed because of the high, variable thermal background, and of the NEAR strategy for background subtraction relying on chopping only. We performed dedicated simulations to verify the feasibility in terms of S/N and we made several adjustments to the QACITS workflow initially designed for $L^{\prime}$-band observations.} Stable pointing control down to 0.015~$\lambda/D$ rms was achieved on timescales of a few hours around meridian passage in good observing conditions. Two features specific to VISIR-NEAR boosted the performance of QACITS with respect to on-sky implementation on instruments operating at shorter IR wavelengths. The first feature is the excellent AO correction provided by the VLT AOF (Strehl ratios $>$95\%), which allows for a good stabilization of the star image behind the AGPM coronagraph. The second feature is the good temporal stability (within $\sim$0.5~pix) of the AGPM center position onto the VISIR detector when tracking a star over several hours during a night. 

QACITS is now part of the NEAR observing template and is routinely used, making coronagraphic observations automatic and stable. In addition to the NEAR campaign, QACITS was also used during the Science Demonstration observations in September and December 2019.

The on-sky validation {for mid-IR observations} of the QACITS centering control approach has important consequences for the development of the mid-IR ELT instrument METIS\cite{Brandl2014}. High-contrast imaging is now a strong science requirement of METIS and several types of coronagraphs have been included in the baseline design, including vortex coronagraphs. Precise centering control of the vortex coronagraphs is mandatory in order to reach the contrast performance required to fulfill the aimed scientific objectives. We demonstrated that QACITS is a robust approach for this purpose.

\subsection*{Disclosures}
The authors have no relevant financial interests and no other potential conflicts of interest to disclose.

\subsection* {Acknowledgments}
The authors thank the ESO Paranal Staff for support in conducting the observations. A.L.M.~acknowledges the financial support of the F.R.S.-FNRS through a postdoctoral researcher grant. O.A.~acknowledges the financial support of the F.R.S.-FNRS through a research associate mandate. This project has received funding from the European Research Council (ERC) under the European Union's Horizon 2020 research and innovation program (grant agreement No 819155) and under the European Union's seventh framework program (grant agreement No 337569). The research was supported by the Wallonia-Brussels Federation (grant for Concerted Research Actions).

\subsection* {Data, Materials, and Code Availability} 
This work is based on observations collected at the European Organisation for Astronomical Research in the Southern Hemisphere under ESO programmes 60.A-9106 and 2102.C-5011.


\bibliography{report}   
\bibliographystyle{spiejour}   



\vspace{1ex}
\noindent Biographies for the authors are not available.


\end{spacing}
\end{document}